\documentclass[letterpaper, 10 pt, conference]{ieeeconf}

\IEEEoverridecommandlockouts

\usepackage{graphicx}
\usepackage{amsmath}
\usepackage{amssymb}
\usepackage{mathtools}
\usepackage{url}
\usepackage{multirow}
\usepackage{booktabs}
\usepackage{multirow}
\usepackage{subcaption}
\usepackage{pifont}

\usepackage{hyperref}

\DeclarePairedDelimiter\norm{\|}{\|}

\newcommand{\f}{\ensuremath\mathbf{f}}
\newcommand{\fl}{\ensuremath\mathbf{f}^\text{L}}
\newcommand{\fnl}{\ensuremath\mathbf{f}^\text{NL}}
\newcommand{\g}{\ensuremath\mathbf{g}}
\newcommand{\gl}{\ensuremath\mathbf{g}^\text{L}}
\newcommand{\gnl}{\ensuremath\mathbf{g}^\text{NL}}
\newcommand{\h}{\ensuremath\mathbf{h}}
\newcommand{\x}{\ensuremath\mathbf{x}}

\newcommand{\R}{\ensuremath\mathbb{R}}
\newcommand{\T}{\ensuremath\mathcal{T}}

\DeclareMathOperator{\cs}{sim}

\DeclarePairedDelimiter{\inner}{\langle}{\rangle}

\title{\LARGE \bf Self-Supervised Feature Learning of 1D Convolutional Neural
  Networks with Contrastive Loss for Eating Detection Using an In-Ear
  Microphone}
\author{Vasileios Papapanagiotou$^{1}$ and Christos Diou$^{2}$ and Anastasios
  Delopoulos$^{1}$%
  \thanks{$^{1}$Vasileios Papapanagiotou and Anastasios Delopoulos are with the
    Multimedia Understanding Group, Dpt. of Electrical and Computer Engineering,
    Faculty of Engineering, Aristotle University of Thessaloniki, Greece {\tt
      vassilis@mug.ee.auth.gr}, {\tt\small adelo@eng.auth.gr}}%
  \thanks{$^{2}$Christos Diou is with Department of Informatics and Telematics,
    Harokopio University of Athens, Greece {\tt\small cdiou@hua.gr}}%
  \thanks{© 2021 IEEE.  Personal use of this material is permitted. Permission
    from IEEE must be obtained for all other uses, in any current or future
    media, including reprinting/republishing this material for advertising or
    promotional purposes, creating new collective works, for resale or
    redistribution to servers or lists, or reuse of any copyrighted component of
    this work in other works.}%
}

\begin{document}

\maketitle

\begin{abstract}
  The importance of automated and objective monitoring of dietary behavior is
  becoming increasingly accepted. The advancements in sensor technology along
  with recent achievements in machine-learning--based signal-processing
  algorithms have enabled the development of dietary monitoring solutions that
  yield highly accurate results. A common bottleneck for developing and training
  machine learning algorithms is obtaining labeled data for training supervised
  algorithms, and in particular ground truth annotations. Manual ground truth
  annotation is laborious, cumbersome, can sometimes introduce errors, and is
  sometimes impossible in free-living data collection. As a result, there is a
  need to decrease the labeled data required for training. Additionally,
  unlabeled data, gathered in-the-wild from existing wearables (such as
  Bluetooth earbuds) can be used to train and fine-tune eating-detection
  models. In this work, we focus on training a feature extractor for audio
  signals captured by an in-ear microphone for the task of eating detection in a
  self-supervised way. We base our approach on the SimCLR method for image
  classification, proposed by Chen et al. from the domain of computer
  vision. Results are promising as our self-supervised method achieves similar
  results to supervised training alternatives, and its overall effectiveness is
  comparable to current state-of-the-art methods. Code is available at
  \url{https://github.com/mug-auth/ssl-chewing}.
\end{abstract}

\section{Introduction}
\label{sec:introduction}

While obesity and eating-related diseases are affecting ever-growing portions of
the population, awareness of our eating habits and behavior can play a very
important role in both prevention and treatment. The under-reporting of eating
in questionnaire-based studies is a well known fact \cite{jessri2016evaluation};
as a result, technology-assisted monitoring using wearable sensors is gaining
more and more attention.

Meaningful information is usually extracted from signals captured by wearable
sensors by means of signal-processing algorithms; these algorithms are often
based on supervised machine learning and thus require labeled training data to
achieve satisfactory effectiveness. This is more important in
deep-learning---based approaches where larger volumes of (annotated) data are
required. Creating such large datasets, however, is challenging. Generating
ground truth annotations requires a lot of manual work, where experts or
specially trained personnel process each part of the dataset in detail in order
to derive the annotations. Besides being laborious and time-consuming, this
process is sometimes prone to errors (which is sometimes reduced by using
multiple annotators for the same data) and can often introduce limitations in
data collection. For example, in the case of video-based annotation, subjects
are limited to the room/area covered by the cameras and data collection cannot
take place in free-living conditions.

One way to overcome this is semi-supervised or unsupervised training
methods. Such methods have already been used with great success in fields such
as speech processing \cite{trigeorgis2016adieu} as well as in applications with
wearable sensors
\cite{nemes2020feature,papadopoulos2020unobtrusive,kyritsis2021}.
Self-supervised methods for image classification have received a lot research
attention recently \cite{xie2020unsupervised}. These methods use multiple
augmentations on unlabeled images in order to learn effective image
representations.

% In the field of computer vision, the use of unlabeled data with multiple
% augmentations in order to improve classifier effectiveness or consistency is
% currently receiving a lot of research interest \cite{xie2020unsupervised}. One
% such recent approach \cite{chen2020simple} uses different augmentations of the
% same image to train a feature-extraction network (based on the ResNet
% architecture) without labels. The training loss is based on a similarity measure
% and aims at achieving high similarity between augmentations of the same original
% image and low similarity between augmentations of different images.

In this work, we adapt ideas from self-supervised image classification to 1D
convolutional neural networks (CNN) with the goal to train a chewing-detection
model on audio from an in-ear microphone, and focus on training the feature
extractor using only unlabeled data. The model is a deep neural network (DNN)
that includes convolutional and max-pooling layers for feature extraction and
fully-connected (FC) layers for classification. We follow the approach of
\cite{chen2020simple,chen2020big} and train the convolutional and max-pooling
layers in a self-supervised way, and then use them as a fixed feature extraction
mechanism to train the FC classification layers. We evaluate on a large and
challenging dataset and compare with our previous work that uses only supervised
training, as well as other algorithms from the literature, and obtain highly
encouraging results.

\section{Network training}
\label{sec:training}

To study if we can successfully train a self-supervised feature extractor we use
the architecture of our previous work on chewing detection
\cite{papapanagiotou2017chewing}. In particular, we focus on the architecture of
$5$ s input window due to its effectiveness in the supervised learning
setting. The network is split in two: the feature extraction sub-network $\f$,
and the classification sub-network, $\h$. Sub-network $\f$ maps an audio window
to a feature vector, i.e.  $\f: \R^{10,000} \rightarrow \R^{512}$ (where
$10,000\, \text{samples} = 5\, \text{s} \cdot 2\, \text{kHz}$) and consists of five
pairs of convolutional layers followed by max-pooling layers. The
convolutional layers have progressively more filters ($8$, $16$, $32$, $64$,
and $64$ respectively) and constant length ($16$ samples for all layers except
for the last one that has $39$); the activation function is ReLU. The
max-pooling ratio is always $2:1$. Sub-network $\g$ classifies a feature vector
to a binary chewing vs. non-chewing decision, i.e.
$\h: \R^{512} \rightarrow [0, 1]$ and consists of two FC layers of $200$ neurons with ReLU
activation followed by a layer of a single neuron with sigmoid activation.

In \cite{papapanagiotou2017chewing}, the entire network\footnote{Operator
  $\circ$ denotes function composition, i.e. $f \circ g (x) = f(g(x))$}
$\h \circ \f$ is trained together on labeled data. In this work, our goal is to
train $\f$ in a self-supervised way, and only use labeled data for training
$\h$.

\subsection{Self-Supervised feature learning}
\label{sec:self-supervised}

To train $\f$ we follow the paradigm of \cite{chen2020simple} where each
training sample $\x$ is augmented with two different augmentations, $\T_{1}$ and
$\T_{2}$, in parallel, yielding samples $\x^{1}$ and $\x^{2}$
respectively. Given an initial training batch of $n$ samples,
$\left\{\x_{i}: i=1, \ldots, n\right\}$ we create a new batch of double the
original size as
$\left\{\x^{1}_{i}: i=1, \ldots, n\right\} \cup \left\{\x^{2}_{i}: i=1, \ldots,
  n\right\}$ using the two augmentations. Given a similarity metric, the network
is then trained to maximize the similarity between all pairs that are derived
from the same original sample, i.e. $\x^{1}_{i}, \x^{2}_{i}$, and to minimize
the similarity between $\x^{k_{1}}_{i}, \x^{k_{2}}_{j}$ for $i \neq j$,
$k_{1}=1,2$, and $k_{2}=1,2$.

Following \cite{chen2020simple}, we use the cosine similarity with temperature
\cite{hinton2015distilling} as our similarity metric:
\begin{equation*}
  % \label{eq:cosine_sim}
  \cs \left( \x_{i}, \x_{j} \right) = \frac
  {\inner{\x_{i}, \x_{j}}}
  {\norm*{\x_{i}} \cdot \norm*{\x_{j}}}
  \tau^{-1}
\end{equation*}
where $\inner{\bullet,\bullet}$ is the inner product, $\norm{\bullet}$ is the Euclidean norm, and
$\tau$ is the temperature parameter. Higher $\tau$ values leads to ``sharper'' softmax
at the network's output, while lower $\tau$ values lead to ``smoother'' output
which can yield more effective representations.

For the contrastive loss function we use the normalized temperature-scaled
cross-entropy loss \cite{oord2019representation} since it has been used on
similar applications, mainly in the domain of 2D signals (such as images). Given
a training batch of $2n$ samples, i.e. $\x^{1}_{i}$ and $\x^{2}_{i}$ for
$i=1,\ldots,n$, the loss for the $i$-th positive pair, i.e. $\x^{1}_{i}$ and
$\x^{2}_{i}$, is defined as:
\begin{equation*}
  % \label{eq:loss_a}
  l_{a}\left(\x^{1}_{i},\x^{2}_{i}\right) = -\ln\frac
  {e^{\cs\left({\x^{1}_{i}, \x^{2}_{i}}\right)}}
  {\sum_{k=1}^{n} \left(\mathbb{I}_{\left[i \neq k\right]}e^{\cs\left(\x^{1}_{i}, \x^{1}_{k}\right)} +
  e^{\cs\left(\x^{1}_{i}, \x^{2}_{k}\right)} \right)}
\end{equation*}
where $\mathbb{I}_{\left[i \neq k\right]} \in \left\{0, 1\right\}$ is a boolean
indicator that is equal to $1$ if and only if $i \neq k$. The indicator is
necessary because the similarity between the a vector and itself is always the
same, i.e. $\cs\left(\x,\x\right)=\tau^{-1}$. This, in turn, renders
$l_{a}\left(\x^{1}_{i},\x^{2}_{i}\right)$ asymmetric, and the final loss between
samples $\x^{1}_{i}$ and $\x^{2}_{i}$ is simply the average:
\begin{equation*}
  % \label{eq:loss}
  l(i) = \frac{1}{2} \left(
  l_{a}\left(\x^{1}_{i},\x^{2}_{i}\right) +
    l_{a}\left(\x^{2}_{i},\x^{1}_{i}\right)
  \right)
\end{equation*}

While it is possible to compute the similarity metric directly on the output of
$\f$, it is usually better to use a projection head, $\g$, that is applied after
$\f$ \cite{chen2020simple}. We experiment with two projection heads: (a) a
linear, $\gl$, that consists of a single FC layer of 128 neurons with linear
activation, and (b) a non-linear, $\gnl$, that consists of 2 FC layers of $512$
neurons with ReLU activations followed by a FC layer of $128$ neurons with
linear activation. Thus, the resulting network is $\g \circ \f$.

To train $\g \circ \f$ we use the LARS optimizer which has been proposed in
\cite{you2017large}. We set the batch size to $256$ (thus obtaining $512$
samples after the dual augmentation process) and train for $100$ epochs, based
on some initial experimentation with the dataset. We apply a warm-up schedule on
the learning rate for $10\%$ of the total epochs (i.e. for $5$ epochs) and reach
a maximum learning rate of $0.3$, after which we apply cosine decay to the
learning rate \cite{gotmare2018closer}.

\subsection{Augmentations}
\label{sec:augmentations}

An important part of this method is selecting augmentations that can help the
training process to learn features that can be effectively used in the final
classification task. Time-stretching (speeding up or slowing down) of audio has
been used both in the speech recognition domain \cite{ko2015audio} as well as in
more general applications (e.g. environmental sounds classification
\cite{salamon2017deep}) with relatively small changes, i.e. $80\%$ to
$120\%$. Other augmentations include pitch shifting, dynamic range compression,
and adding background noise \cite{salamon2017deep}.

Based on the above as well as the nature of audio chewing signals, we focus on
two augmentations: global amplification level and background noise. In
particular, global amplification is $\T_{1}\left(\x\right) = \alpha \cdot \x$
where $\alpha$ is the global amplification level and is drawn from a uniform
distribution in the range $[0.5, 2.0]$ (a different $\alpha$ is drawn for each
$\x$). This choice for this augmentation is based on our experience with in-ear
microphone signals \cite{boer2018splendid}, where placement, ear and sensor
shape compatibility, and even movement can change the overall amplification
level of the captured audio.

The second augmentation is the addition of noise and implemented as:
$\T_{2}\left(\x\right) = \x + \mathbf{v}$ where $\mathbf{v}$ is a vector of IIR
``noise'' samples, drawn from uniform distributions in the range of
$[-0.005, 0.005]$. Value $0.005$ has been chosen as it roughly equal to $0.1$ of
the average standard deviation of audio signal across our entire dataset,
yielding $20$ dB SNR. Adding noise to the audio signal simulates noisy
environments (such as noisy city streets, restaurants, etc) and helps our
feature extractor learn to ``ignore'' its influence.

\subsection{Supervised classifier training}
\label{sec:supervised}

Given the trained feature extractor $\f$ we can now train a chewing detection
classifier based on label data. In this stage, the projection head $\g$ can be
discarded and thus the final network is $\h \circ \f$, where only the weights of
$\h$ are trained. It is possible, however, to retain a part of the projection
head in the final model \cite{chen2020big}. We do this for the case of the
non-linear projection head, $\gnl$: let $\gnl=\gnl_{2} \circ \gnl_{1}$ where
$\gnl_{1}$ corresponds to the first layer of $\gnl$ and $\gnl_{2}$ to the
remaining (second and third) layers of $\gnl$; thus, the final network is
$\h \circ \gnl_{1} \circ \f$ (again, only weights of $\h$ are trained here).

We use the ADAM optimizer \cite{kingma2017adam} with a learning rate of
$10^{-3}$ and minimize binary cross-entropy based on ground truth values:
\begin{equation*}
  \label{eq:binary_xentropy}
  H\left(\hat y_{i}; y_{i}\right) =
  -y_{i} \log \hat{y}_{i} -(1-y_{i}) \log \left(1-\hat{y}_{i}\right)
\end{equation*}
where $y_{i} \in \left\{0, 1\right\}$ is the ground truth value for the $i$-th
sample and $\hat{y} \in [0, 1]$ is the output of $\h$.

\subsection{Post-processing of predicted labels}
\label{sec:post_processing}

The predicted labels indicate chewing vs. non-chewing; thus, chewing ``pulses''
correspond to individual chews. Similarly to our previous works
\cite{papapanagiotou2017novel,papapanagiotou2017chewing}, we aggregate chews to
chewing bouts and then chewing bouts to meals. In short, (a) chewing bouts are
obtained by merging chews that are no more than $2$ s apart, (b) chewing bouts
of less than $5$ s are discarded, (c) meals are obtained by merging chewing
bouts that are no more than $60$ s apart, (d) meals for which the ratio of
``duration of bouts'' over ``duration of meal'' is less than $25\%$ are
discarded.

\section{Dataset}
\label{sec:dataset}

The dataset we use has been collected in the Wageningen University in 2015
during a pilot study of the EU SPLENDID project \cite{maramis2014}. Recordings
from $14$ individuals (approximately $60$ h) are available. Each subject had two
meals in the university premises and was free to leave the university, engage in
physical activities, and have as many other meals and snacks wished for the rest
of the recording time. This dataset has also been used in
\cite{papapanagiotou2017novel} and \cite{papapanagiotou2017chewing}.

The sensor is a prototype in-ear microphone sensor consisting of Knowles
FG-23329-D65 microphone housed in a commercial ear bud. Audio was originally
captured at $48$ kHz but we have down-sampled it at $2$ kHz (as in
\cite{papapanagiotou2017novel}) to reduce the computational burden; we have also
applied a high-pass Butterworth filter with a cut-off frequency of $20$ Hz to
remove very low spectrum content and the effect of DC drifting that was present
in the chewing-sensor prototype.

\section{Evaluation}
\label{sec:evaluation}

We split our dataset of $14$ subjects into two parts: a ``development'' set with
$10$ subjects (selected randomly), $S_{1}$, and a ``final evaluation'' set with
the remaining $4$ subjects, $S_{2}$. Note that $S_{1}$ and $S_{2}$ are
disjoint. In the first part of the evaluation, we explore training
hyper-parameters and architecture choices on $S_{1}$. In the second part, we
apply what we learned and, after training our models on $S_{1}$, we evaluate
them on $S_{2}$.

In the first part of evaluation, our goal is to understand the effect of the
temperature and the projection head on classification accuracy. In particular,
we first train $\g\circ\f$ on the entire $S_{1}$ (all $10$ subjects) using
self-supervised training (as described in Section \ref{sec:self-supervised}). We
then train $\h$ on the same $10$ subjects but in a supervised way (as described
in Section \ref{sec:supervised}) in typical leave-one-subject-out (LOSO)
fashion. During each LOSO iteration, data from $9$ subjects are available for
training. We select a small part of the $9$ subjects (specifically $2$ subjects)
as a validation set, and train on the remaining ($7$ subjects). We train for
$100$ epochs and compute the loss over the validation subjects' data after each
epoch; we select the model that minimizes the validation loss at the end of each
epoch. We use a batch size of $64$. Note that in these experiments
self-supervised feature learning takes place in all $10$ subjects of $S_{1}$
(for computational reasons) and are only used here to obtain an assessment of
the effect of temperature and not for evaluating our algorithm. Results of
evaluation on the held-out dataset ($S_{2}$) are presented in Table
\ref{tab:comparison}.

We examine different values of temperature $\tau$; results are presented in Tables
\ref{tab:results1} - \ref{tab:results3}. Table \ref{tab:results1} shows results
for the $\h \circ \fl$, where $\fl$ is $\f$ trained with the linear projection head
$\gl$. Based on F1-score, the best results are obtained for $\tau=0.5$ while most
other values of $\tau$ yield F1-score higher then $0.7$.

Table \ref{tab:results2} shows similar results; in this case the network is
$\h \circ \fnl$, where $\fnl$ is $\f$ trained with the non-linear projection head
$\gnl$. Highest F1-score is obtained for $\tau=0.1$; in general, training seems to
benefit more from smaller temperature values. Extremely large temperature values
(e.g. $100$) seem to not be beneficial (this is also observed in
\cite{chen2020big}).

Finally, Table \ref{tab:results3} shows similar results for
$\h \circ \gnl_{1} \circ \fnl$; this network is the same as before (i.e.
$\h \circ \fnl$) but the first layer of the projection head, $\gnl_{1}$, is
retained. Here, smaller temperatures seems to benefit the overall effectiveness
more, with $\tau=1$ yielding the highest F1-score, and high temperatures
($\tau \geq 50$) degrade the effectiveness completely.

\begin{table}
  \centering
  \caption{Results for the $\h \circ \fl$ network on $S_{1}$.}
  \label{tab:results1}
  \begin{tabular}{cccccc}
    \toprule
    $\tau$ & \textbf{prec.} & \textbf{rec.} & \textbf{F1-score} & \textbf{acc.} & \textbf{w. acc.}\\
    \midrule
    $0.1$ & $0.82$ & $0.64$ & $0.72$ & $0.94$ & $0.84$ \\
    $0.5$ & $0.75$ & $0.76$ & $0.76$ & $0.94$ & $0.86$ \\
    $1$   & $0.82$ & $0.55$ & $0.66$ & $0.93$ & $0.81$ \\
    $5$   & $0.82$ & $0.62$ & $0.70$ & $0.93$ & $0.83$ \\
    $10$  & $0.80$ & $0.66$ & $0.72$ & $0.94$ & $0.84$ \\
    $50$  & $0.77$ & $0.12$ & $0.21$ & $0.88$ & $0.57$ \\
    $100$ & $0.85$ & $0.63$ & $0.72$ & $0.94$ & $0.84$ \\
    \bottomrule
  \end{tabular}
\end{table}

\begin{table}
  \centering
  \caption{Results for the $\h \circ \fnl$ network on $S_{1}$.}
  \label{tab:results2}
  \begin{tabular}{cccccc}
    \toprule
    $\tau$ & \textbf{prec.} & \textbf{rec.} & \textbf{F1-score} & \textbf{acc.} & \textbf{w. acc.}\\
    \midrule
    $0.1$ & $0.85$ & $0.67$ & $0.75$ & $0.94$ & $0.86$ \\
    $0.5$ & $0.85$ & $0.63$ & $0.72$ & $0.94$ & $0.85$ \\
    $1$   & $0.85$ & $0.63$ & $0.73$ & $0.94$ & $0.85$ \\
    $5$   & $0.83$ & $0.50$ & $0.62$ & $0.92$ & $0.79$ \\
    $10$  & $0.84$ & $0.58$ & $0.69$ & $0.93$ & $0.82$ \\
    $50$  & $0.76$ & $0.59$ & $0.66$ & $0.92$ & $0.81$ \\
    $100$ & $0.86$ & $0.31$ & $0.45$ & $0.91$ & $0.70$ \\
    \bottomrule
  \end{tabular}
\end{table}

\begin{table}
  \centering
  \caption{Results for the $\h \circ \gnl_{1} \circ \fnl$ network on $S_{1}$.}
  \label{tab:results3}
  \begin{tabular}{cccccc}
    \toprule
    $\tau$ & \textbf{prec.} & \textbf{rec.} & \textbf{F1-score} & \textbf{acc.} & \textbf{w. acc.}\\
    \midrule
    $0.1$ & $0.75$ & $0.72$ & $0.73$ & $0.93$ & $0.85$ \\
    $0.5$ & $0.74$ & $0.59$ & $0.66$ & $0.92$ & $0.81$ \\
    $1$   & $0.87$ & $0.64$ & $0.74$ & $0.94$ & $0.85$ \\
    $5$   & $0.80$ & $0.57$ & $0.67$ & $0.93$ & $0.81$ \\
    $10$  & $0.88$ & $0.02$ & $0.03$ & $0.88$ & $0.48$ \\
    $50$  & $0.73$ & $0.32$ & $0.44$ & $0.90$ & $0.69$ \\
    $100$ & $0.75$ & $0.24$ & $0.36$ & $0.89$ & $0.65$ \\
    \bottomrule
  \end{tabular}
\end{table}

In the second part of the evaluation, we use evaluate on the $4$ subjects of
$S_{2}$ with models trained on $S_{1}$. In particular, we select the best
network of the three different approaches (Tables \ref{tab:results1} -
\ref{tab:results3}) based on F1-score. We train $\f$ and $\g$ on $S_{1}$ in an
self-supervised way and then train $\h$ again on $S_{1}$ in a supervised
way. The three trained models are then evaluated on $S_{2}$ and the results are
shown in Table \ref{tab:comparison}. To compare, we also train a fourth model by
training the entire network (i.e. $\h \circ \f$) on $S_{1}$ in a supervised way
(similar to how models are trained in \cite{papapanagiotou2017chewing}). Results
are shown in the fourth line of Table \ref{tab:comparison}.

Results are particularly encouraging as the three networks with the
self-supervised trained feature-extraction layers (lines $1$-$3$) not only
achieve similar effectiveness with the completely supervised-trained network
(line $4$) but also improve over it. The non-linear projection head (without
retainment of the first layer, i.e. the $\h \circ \fnl$) achieves the highest
F1-score and weighed accuracy among all the three self-supervised models and the
supervised model.

\begin{table}
  \centering
  \caption{Final evaluation of effectiveness on $S_{2}$, when training on
    $S_{1}$ between self-supervised feature learning and supervised classifier
    training (lines 1-3), and supervised training of the entire network (line
    4).}
  \label{tab:comparison}
  \begin{tabular}{lcccccc}
    \toprule
    \textbf{model} & $\tau$ & \textbf{prec.} & \textbf{rec.} & \textbf{F1-score} & \textbf{acc.} & \textbf{w. acc.}\\
    \midrule
    \midrule
    $\h \circ \fl$             & $0.5$ & $0.83$ & $0.79$ & $0.81$ & $0.95$ & $0.89$ \\
    $\h \circ \fnl$            & $0.1$ & $0.94$ & $0.79$ & $0.86$ & $0.97$ & $0.92$ \\
    $\h \circ \gnl_{1} \circ \fnl$ & $1$   & $0.84$ & $0.78$ & $0.81$ & $0.95$ & $0.89$ \\
    \midrule
    $\h \circ \f$ & -         & $0.93$ & $0.61$ & $0.74$ & $0.95$ & $0.84$ \\
    \bottomrule
  \end{tabular}
\end{table}

As a final comparison, we repeat the results of three different algorithms of
\cite{passler2012evaluation} and for the $5$-sec architecture of
\cite{papapanagiotou2017chewing}, as presented in
\cite{papapanagiotou2017chewing}. It is important to note that these results are
averages across all $14$ subjects of the dataset, so they are not directly
comparable with all previous results. However, they can give an estimate about
the overall effectiveness of our approach. Results indicate that self-supervised
training exceeds the effectiveness of several of the methods proposed in the
bibliography. Only the $5$-sec architecture achieves better results (F1-score of
$0.91$ versus $0.84$ of $\h \circ \gnl_{1} \circ \fnl$).

\begin{table}
  \centering
  \caption{Results for comparison with base-line as presented in
    \cite{papapanagiotou2017chewing}: three algorithms of
    \cite{passler2012evaluation} and the $5$-sec arch. CNN chewing detector of
    \cite{papapanagiotou2017chewing} (supervised LOSO on all $14$ subjects,
    batch size of $16$, Adam optimizer with learning rate of $0.0016$, and
    $5 \cdot 10^5$ epochs).}
  \label{tab:results_soa}
  \begin{tabular}{lccccc}
    \toprule
    \textbf{approach} & \textbf{prec.} & \textbf{rec.} & \textbf{F1-score} & \textbf{acc.} & \textbf{w. acc.}\\
    \midrule
    MSEA \cite{passler2012evaluation}  & $0.29$ & $0.80$ & $0.42$ & $0.72$ & $0.76$ \\
    MESA \cite{passler2012evaluation}  & $0.30$ & $0.81$ & $0.44$ & $0.74$ & $0.77$ \\
    LPFSA \cite{passler2012evaluation} & $0.29$ & $0.81$ & $0.43$ & $0.72$ & $0.76$ \\
    $5$-sec arch. \cite{papapanagiotou2017chewing} & $0.89$ & $0.93$ & $0.91$ & $0.98$ & $0.96$ \\
    \bottomrule
  \end{tabular}
\end{table}

\section{Conclusions}
\label{sec:conclusions}

In this work, we have presented an approach for training the feature extraction
layers of an audio-based chewing-detection neural network in an self-supervised
way. Self-supervised training seems to lead to highly effective models, while at
the same time reducing the labor needed for manual annotation as well as takes
advantage of large amounts of unlabeled data for representation learning. Our
experiments show very promising results, as self-supervised training achieves
similar, and sometimes better, effectiveness compared to a similar
fully-supervised approach. Additionally, best results (F1-score of $0.86$) are
comparable to fully-supervised methods on the same dataset (F1-score of $0.91$
of \cite{papapanagiotou2017chewing}). Future work includes further studying the
effect of additional augmentations for the self-supervised training part, and
examining how much effectiveness is affected when there are less data available
for the supervised training part. Additionally, the effectiveness of the trained
self-supervised network can be evaluated on other problems such as individual
chew detection or food type recognition.

\section*{Acknowledgments}

The work leading to these results has received funding from the EU Commission
under Grant Agreement No. 965231, the REBECCA project (H2020).

\bibliographystyle{IEEEtran}
\bibliography{IEEEabrv, root}

\end{document}